\documentclass[showpacs,reprint,amsmath,amssymb,aps,prl]{revtex4-1}
\usepackage{amsmath,amssymb,graphicx}

\newcommand{\E}{\mathrm{e}}
\newcommand{\I}{\mathrm{i}}

\begin{document}

\title{Ptychographic ultrafast pulse reconstruction}

\author{D. Spangenberg}
\author{E. Rohwer}
\affiliation{LRI, Stellenbosch University, Private Bag X1, 7602 Matieland, South Africa}

\author{M.H. Br\"ugmann}
\author{T. Feurer}\email{Corresponding author: thomas.feurer@iap.unibe.ch}
\affiliation{IAP, University of Bern, Sidlerstr. 5, 3012 Bern, Switzerland}

\begin{abstract}
We demonstrate a new ultrafast pulse reconstruction modality which is somewhat reminiscent of frequency resolved optical gating but uses a modified setup and a conceptually different reconstruction algorithm that is derived from ptychography. Even though it is a second order correlation scheme it shows no time ambiguity. Moreover, the number of spectra to record is considerably smaller than in most other related schemes which, together with a robust algorithm, leads to extremely fast convergence of the reconstruction.
\end{abstract}


\maketitle 


Since the discovery of mode-locking and femtosecond laser pulses, characterization of these pulses has been an imminent task as no known detector was, and still is, fast enough to measure such ultrashort pulses directly. In 1967 H.~Weber showed that an auto-correlation scheme together with a nonlinear process can be employed to estimate the pulse duration \cite{Weber1967}, such arrangement is now classified as a second order background-free intensity auto-correlation measurement. About 20 years later, J.-C.~Diels and coworkers introduced the second order interferometric auto-correlation with the aim to not only estimate the pulse duration but to fully characterize the electric field of ultrashort laser pulses, e.g. their spectral amplitude and phase \cite{Diels_et_al1985}. In the following years several further ingenious schemes were proposed and demonstrated, such as FROG \cite{Trebino2000}, STRUT \cite{Chilla1991}, SPIDER \cite{Iaconis1998}, PICASO \cite{Nicholson_et_al1999}, MIIPS \cite{Xu_et_al2006} and many variations thereof. Often these modalities were first demonstrated at optical wavelengths but were later transferred to other spectral regions ranging from the MIR to the X-ray regime.

Here, we present a new modality for ultrafast electric field characterization which is a modification of a phase retrieval scheme widely used in lens-less imaging, namely ptychography. It is related to the solution of the phase problem in crystallography as proposed by Hoppe \cite{Hoppe1969} and was first demonstrated at optical wavelengths \cite{Rodenburg_et_al2007}. In ptychography the real space image of an object, in particular its amplitude and phase, is reconstructed iteratively from a series of far-field diffraction measurements. Each of those is recorded after either moving the object or the coherent illumination beam in a plane perpendicular to the propagation direction of the illumination beam. The transverse shift of the illumination beam is smaller than its spatial support, so that subsequent far-field diffraction patterns result from different, but overlapping regions of the real space object. The spatial resolution is limited by the positioning accuracy, the stability of the entire setup, but mainly by the angular range of scattered wavevectors that can be recorded with a sufficiently high signal-to-noise ratio. Ptychography has been proven to produce the correct real space image if the illumination beam is known \cite{McCallumRodenburg1992}, but seems to work even if the illumination beam is unknown. In this case its profile is reconstructed together with the real space image \cite{Thibault_et_al2009, MaidenRodenburg2009}.

\begin{figure}[htb]
\includegraphics[width=80mm]{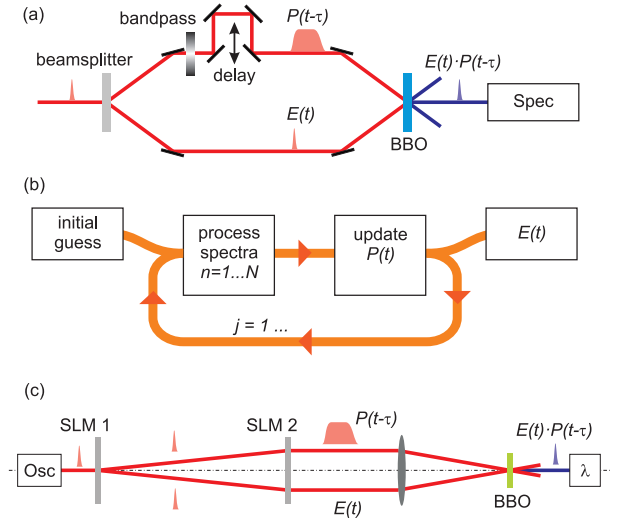}
\caption{\label{fig1} a) Schematic of a generic setup, b) flowchart of the algorithm, and c) schematic of the setup used here.}
\end{figure}

Applying ptychography to reconstructing temporal rather than spatial objects, requires operating in one-dimensional space with the conjugated variables time and frequency. In general, the one-dimensional phase retrieval problem is ambiguous and different solutions may result in the same spectra. In the framework of ptychography, however, the uniqueness of the solution in two \cite{McCallumRodenburg1992} but also one dimension \cite{Guizar-Sicairos2010} is warranted as long as the illumination pulse is known. In analogy to the spatial problem, the unknown temporal object is to be reconstructed iteratively from a series of far-field diffraction measurements, i.e. spectra. Each of those is recorded after delaying the coherent illumination pulse with respect to the temporal object with the time delay being smaller than the temporal support of the illumination pulse. That is, subsequent spectra result from different but overlapping parts of the unknown temporal object. The temporal resolution is primarily limited by the range of spectral amplitudes which can be recorded with a sufficiently high signal-to-noise ratio. Recently, we have shown that ptychography can indeed be applied to reconstruct temporal objects if the illumination pulse is fully characterized \cite{Spangenberg_et_al2014a}. Here, we show that the concept can be extended to ultrafast pulse characterization where the illumination pulse is in principle unknown but derived from the also unknown input pulse. 


The basic scheme is illustrated in Fig.~\ref{fig1}a). The unknown input pulse is divided in two replicas by a beam splitter. The reflected replica, i.e. the temporal object $E(t)$, propagates via mirrors to the nonlinear crystal. The transmitted replica, i.e. the illumination pulse $P(t)$, passes through a spectral bandpass filter, is delayed in time, and overlapped with $E(t)$ in the nonlinear crystal. The resulting sum-frequency light, i.e. $E(t) P(t-\tau)$, is analyzed by a spectrometer. While the setup is similar to a second harmonic FROG \cite{Trebino2000}, except for the spectral bandpass, the data processing and the phase reconstruction algorithm is conceptually very different. The reconstruction algorithm, which is derived from the ptychographic iterative engine \cite{Faulkner2005}, is schematically depicted in Fig.~\ref{fig1}b). Its task is to determine the unknown electric field $E(t)$ of the input pulse from a sequence of $N$ sum-frequency spectra $S_n(\omega)$ recorded at different time delays $\tau_n$ between $E(t)$ and $P(t)$, with $n = 1 \ldots N$. As a starting point for the reconstruction algorithm we assume white noise for the electric field, i.e. $E_{j=1,n=0}(t)$, and the time delayed inverse Fourier transforms of the spectral bandpass filter $T(\omega)$ for the initial illumination pulses $P_{j=1,n}(t-\tau_n)$. In every iteration $j$ all measured spectra $S_n(\omega)$ are processed. For ascending $n$ the algorithm calculates the exit field $\xi_{j,n}(t)$ for a particular time delay $\tau_n$ between the illumination pulse $P_{j,n}(t-\tau_n)$ and the current estimate of the pulse's electric field $E_{j,n-1}(t)$

\begin{equation}
\xi_{j,n}(t) = E_{j,n-1}(t) \; P_{j,n}(t-\tau_n).
\end{equation}

From $\xi_{j,n}(t)$ we calculate the Fourier transform $\xi_{j,n}(\omega)$ and replace its modulus by the square root of the corresponding spectrum $S_n(\omega)$ while preserving its phase. After an inverse Fourier transformation the new function $\xi'_{j,n}(t)$ differs from the initial estimate and the difference is used to update the current estimate of the electric field

\begin{eqnarray}
\label{eq_pty_iter}
\nonumber
E_{j,n}(t) & = & E_{j,n-1}(t) + \beta \; U_{j,n}(t-\tau_n) \\
&& \times [\xi'_{j,n}(t) - \xi_{j,n}(t)]
\end{eqnarray}

with the weight or window function

\begin{equation}
\label{eq_pty_wt}
U_{j,n}(t) = \frac{|P_{j,n}(t)|}{\mathrm{max}(|P_{j,n}(t)|)} \; \frac{P_{j,n}^*(t)}{|P_{j,n}(t)|^2+\alpha}
\end{equation}

and the two constants $\alpha < 1$ and $\beta \in \; ]0 \ldots 1]$. Before proceeding with the next iteration $j+1$ we set $E_{j+1,n=0}(t) = E_{j,n=N}(t)$ and update the illumination pulse according to

\begin{equation}
\label{eq_Pt}
P_{j+1}(t-\tau_n) = \mathcal{F}^{-1}\left[ E_{j,n=N}(\omega) \; T(\omega) \; \E^{\I (\omega-\omega_0) \tau_n} \right].
\end{equation}

Updating $P_{j,n}(t)$ is not part of the original ptychographic reconstruction algorithm but is crucial here to achieve convergence. The best approximation to the actual input pulse appears typically after only a few iterations $j$.

The ptychographic scheme is different from other phase reconstruction modalities, such as principal components generalized projection \cite{Kane2008} or blind deconvolution \cite{DeLong1995}. 1) Through $\alpha$ the update function can be adjusted for different signal-to-noise levels. 2) There is no time ambiguity in the measurement, because the bandpass filter breaks the symmetry. 3) The time delay increment is not related to the desired temporal resolution or the wavelength sampling of the spectrometer, but only to the duration of $P(t)$. The time delays must not even be equidistant. 4) Typically, only a few spectra have to be recorded. For example, assume an illumination pulse of 1~ps duration and a temporal support of $E(t)$ of 0.5~ps; with a 10\% overlap between successive measurements, three spectra recorded for time delays of $-0.9$~ps, $0$~ps, and $0.9$~ps are sufficient to reconstruct $E(t)$. 5) The small number of spectra to process and the robust algorithm result in an extremely fast convergence of the retrieval algorithm.


Before performing experiments we ran simulations with the aim to identify the optimal experimental parameters, i.e. the bandpass filter, the time delays $\tau_n$, and the number of spectra $N$ to record, as well as the optimal reconstruction parameters, i.e. $\alpha$ and $\beta$. We found that the bandpass filter should be centered at the laser's baseband frequency and its width should be less than half of the laser's bandwidth. Smaller widths yield faster convergence, however, at the expense of signal strength. The spectral width of the bandpass then determines the full width at half maximum of the illumination pulse, i.e. $\mathrm{FWHM}\{P(t)\}$. The time delay increment should be approximately $0.5 \cdot \mathrm{FWHM}\{P(t)\}$. At this point it is important to note that the time delay is related to the temporal duration of the illumination pulse and not determined by the sampling of the spectrogram, as it is the case for example in FROG. Conversely, the temporal resolution is determined by the largest spectral sidebands which can be detected with a sufficient SNR. In the reconstruction the choice of $\alpha$ is determined mostly by the signal-to-noise ratio (SNR) and $\beta$ by the illumination pulse duration and the time delay. A suitable value for $\alpha$ was determined by analyzing the rms error (rms refers to the root mean square difference between theoretical and reconstructed spectrogram) for sets of simulated noisy spectra as a function of $\alpha \in \; [0 \ldots 1]$ and the SNR. For a measured SNR of $>500$, $\alpha \approx 0.2$ leads to the smallest rms. Similarly, the value for $\beta \approx 0.5$ was determined by analyzing the rms error as function of $\beta \in \; ]0 \ldots 1]$ and the ratio of the time delay increment and the illumination pulse duration. We would like to emphasize that $\alpha$ as well as $\beta$ can be varied within a wide range of values, i.e. $\alpha = 0.05 \ldots 0.5$ and $\beta = 0.2 \ldots 0.9$, the final result does not change, however, the rate of convergence may slow down. 


In order to vary all parameters in the most flexible way we demonstrate the principle with a setup that allows for individually tailoring both replicas of the incident pulse. A schematic of the experimental setup is shown in Fig.~\ref{fig1}c). The pulse source is an 80~MHz Ti:sapphire oscillator which delivers 80~fs pulses centered at 800~nm. The first two-dimensional spatial light modulator (SLM1) is loaded with a binary hologram to diffract the incoming beam to the plus and the minus first diffraction orders and replaces the beam splitter in Fig.~\ref{fig1}a). Both replicas can be independently modulated by a pulse shaping apparatus~\cite{Vaughan2004}, which includes a second two-dimensional spatial light modulator (SLM2), and are subsequently focused to a 100~$\mu$m thick beta-bariumborate crystal where the exit field, i.e. the product field, is produced through sum-frequency generation. The resulting spectra are centered at 400~nm and are analyzed by a spectrometer covering the range of 300~nm to 545~nm with a resolution of 0.18~nm. That is, the spectrometer range is approximately $\pm 100$~nm around the center wavelength (400~nm) which yields a minimum theoretical temporal resolution of the measurement apparatus of approximately 5~fs. For FROG measurements the same phase modulation is applied to both replicas. One of them is time delayed in increments of 20~fs and a total of 301 spectra are recorded. In ptychography we also imprint the same phase modulation to both replicas but in addition we spectrally filter one of them. The bandpass is programmed to have a rectangular transmission characteristic with a width of 3~nm centered at 800~nm. The filtered replica is time delayed in increments of 300~fs and 11~spectra are recorded.


\begin{figure}[htb]
\includegraphics[width=80mm]{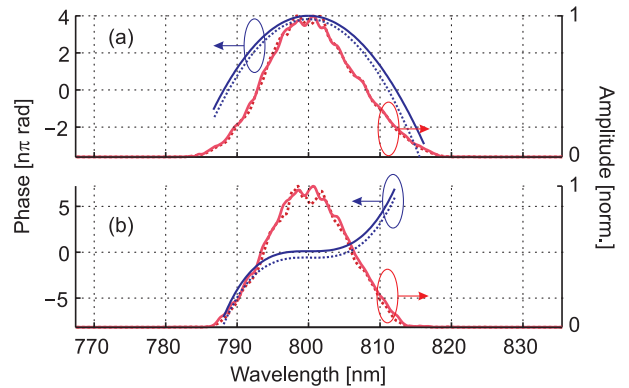}
\caption{\label{fig2} Spectral amplitude (red curves) and phase (blue curves) for a) a second order phase of $\phi_2=-10^4$~fs$^2$ and b) a third order phase of $\phi_3=5 \cdot 10^5$~fs$^3$. The solid curves represent the measured spectral amplitudes and the programmed phases and the dashed curves the retrieved amplitudes and phases.}
\end{figure}

Figure~\ref{fig2} shows the retrieved spectral amplitude (red curves) and phase (blue curves) for a second order phase of $\phi_2=-10^4$~fs$^2$ and a third order phase of $\phi_3=5 \cdot 10^5$~fs$^3$, respectively. Ptychography recovers both the absolute value and the sign of the spectral phase correctly and the agreement with the measured amplitude and the programmed phase is good. Note that a small absolute phase offset was introduced between the theoretical and the reconstructed phases for better comparison. Since the scheme is somewhat reminiscent to a SHG FROG arrangement, we compare our results in the following to those obtained by an SHG FROG. 

\begin{table}[htb]
\caption{\label{tab1} Columns 1 to 3 show the programmed second, third, and fourth order phases. The numbers given ($-1$, 0 , 1) have to be multiplied by $10^4$~fs$^2$, $2.5 \cdot 10^5$~fs$^3$, and $5 \cdot 10^6$~fs$^4$ for the second, third, and fourth order phase, respectively. Columns 4 to 6 show results of the FROG retrieval and columns 7 to 9 results from the ptychographic scheme.}
\begin{tabular}{rrr|rrr|rrr}
\hline
\multicolumn{3}{c|}{target} & \multicolumn{3}{c|}{FROG} & \multicolumn{3}{c}{Ptychography} \\
$\phi_2$ & $\phi_3$ & $\phi_4$ & $\phi_2$ & $\phi_3$ & $\phi_4$ & $\phi_2$ & $\phi_3$ & $\phi_4$ \\
\hline
 0 &  1 &  0 &  0.05 &  1.18 & -0.18 &  0.05 &  1.15 & -0.13 \\
 0 &  0 &  1 &  0.01 & -0.09 &  0.97 &  0.02 &  0.07 &  0.84 \\
 1 &  0 & -1 &  0.93 &  0.04 & -0.90 &  0.90 & -0.01 & -0.90 \\
 1 &  1 & -1 &  0.81 &  0.83 & -0.85 &  0.94 &  1.04 & -0.75 \\
-1 & -1 &  1 & -0.90 & -1.12 &  1.09 & -1.03 & -1.01 &  0.88 \\
 1 & -1 & -1 &  1.08 & -0.83 & -0.93 &  1.05 & -0.94 & -1.03 \\
\hline
\end{tabular}
\end{table}

\begin{figure}[htb]
\includegraphics[width=80mm]{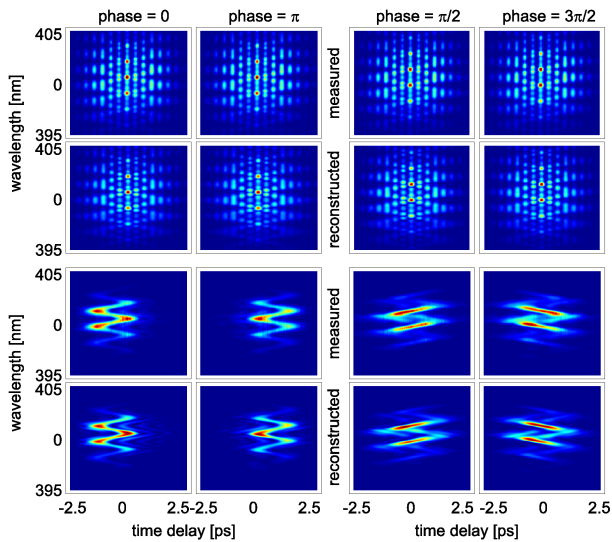}
\caption{\label{fig3} The measured (first row) and the reconstructed (second row) FROG spectrograms and the measured (third row) and the reconstructed (fourth row) ptychographic spectrograms. From the first to the fourth column the absolute phase changes, i.e. $\varphi = 0, \pi, \pi/2, 3\pi/2$.}
\end{figure}

Next, we analyze different combinations of second, third, and fourth order phase contributions and the results of the ptychographic and the FROG retrievals are summarized in table~\ref{fig3}. The second, third, and fourth order phases were $\phi_2 = \pm 10^4$~fs$^2$, $\phi_3 = \pm 2.5 \cdot 10^5$~fs$^3$, and $\phi_4 = \pm 5 \cdot 10^6$~fs$^4$, respectively. When disregarding the sign issue, we find that both, FROG and ptychography, produce correct values for the six different combinations of $\phi_{1,2,3}$ tested. Lastly, we investigated a sinusoidal phase modulation, i.e. $\phi(\omega) = A \sin[(\omega-\omega_0) T + \varphi]$. Figure~\ref{fig3} shows the results for $A=2.41$, $T=350$~fs, and $\varphi = 0, \pi/2, \pi, 3\pi/2$. The top row shows the measured and the second row the reconstructed FROG spectrograms. The third and the fourth row show the measured and the reconstructed ptychographic spectrograms. For both modalities, we find excellent agreement between the measured and the reconstructed spectrograms. Note, that only a small subset of spectra displayed here, i.e. 11 out of 301, is used for the ptychographic reconstruction.

\begin{figure}[htb]
\includegraphics[width=70mm]{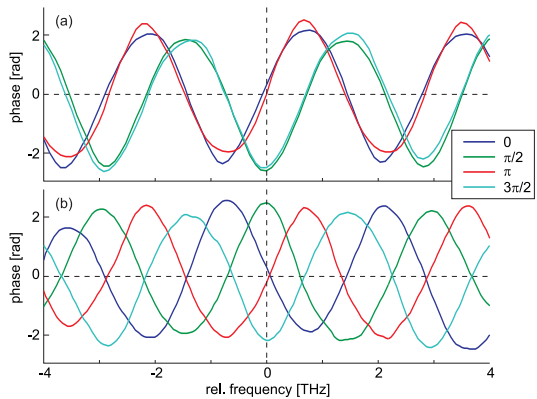}
\caption{\label{fig4} Oscillatory phases reconstructed from a) FROG and b) through ptychography.}
\end{figure}

Fig.~\ref{fig4} shows the reconstructed phases for (a) FROG and (b) ptychography for the data in Fig.~\ref{fig3}. While FROG can distinguish between a sinusoidal ($\varphi=0$) and a co-sinusoidal ($\varphi=\pi/2$) phase oscillation but not between two sin(cosin)usoidal phase oscillations with opposite sign, i.e. $\varphi=0,\pi$ ($\varphi=\pi/2, 3\pi/2$), ptychography can differentiate between all four absolute phases.

In conclusion, we have demonstrated a new modality for ultrashort pulse reconstruction based on a retrieval algorithm derived from ptychography which --- despite being second order --- shows no time ambiguity, requires to record only a small number of spectra, converges extremely fast and reliable,
and can be easily extended to other nonlinear processes.


\begin{acknowledgments}
We gratefully acknowledge fruitful discussions with P.~Neethling and financial support from the National Research Foundation, the CSIR NLC, and the NCCR MUST research instrument of the Swiss National Science Foundation.
\end{acknowledgments}



\begin{thebibliography}{15}
\expandafter\ifx\csname natexlab\endcsname\relax\def\natexlab#1{#1}\fi
\expandafter\ifx\csname bibnamefont\endcsname\relax
  \def\bibnamefont#1{#1}\fi
\expandafter\ifx\csname bibfnamefont\endcsname\relax
  \def\bibfnamefont#1{#1}\fi
\expandafter\ifx\csname url\endcsname\relax
  \def\url#1{\texttt{#1}}\fi
\expandafter\ifx\csname urlprefix\endcsname\relax\def\urlprefix{URL }\fi
\providecommand*{\bibinfo}[2]{#2}
\providecommand*{\eprint}[1]{#1}
\providecommand*{\url}[1]{#1}
\begingroup\makeatletter
 \@temptokena{%
  \expandafter\ifx\csname citenamefont\endcsname\relax
   \DeclareRobustCommand\citenamefont{\@firstofone}%
   \global\let\citenamefont\citenamefont
   \global\expandafter\let\csname citenamefont \expandafter\endcsname\csname
  citenamefont \endcsname
  \fi
 }\if@filesw\immediate\write\@auxout{\the\@temptokena}\fi
\expandafter\endgroup\the\@temptokena

\bibitem[{\citenamefont{Weber}(1967)}]{Weber1967}
\bibinfo{author}{\bibfnamefont{H.~P.} \bibnamefont{Weber}},
  \bibinfo{journal}{Journal of Applied Physics}
  \textbf{\bibinfo{volume}{38}}(\bibinfo{number}{5}), \bibinfo{pages}{2231}
	(\bibinfo{year}{1967}).

\bibitem[{\citenamefont{Diels} \emph{et~al.}(1985)\citenamefont{Diels, Dietel,
  Fontaine, Rudolph, and Wilhelmi}}]{Diels_et_al1985}
\bibinfo{author}{\bibfnamefont{J.-C.} \bibnamefont{Diels}},
  \bibinfo{author}{\bibfnamefont{W.}~\bibnamefont{Dietel}},
  \bibinfo{author}{\bibfnamefont{J.~J.} \bibnamefont{Fontaine}},
  \bibinfo{author}{\bibfnamefont{W.}~\bibnamefont{Rudolph}}, \bibnamefont{and}
  \bibinfo{author}{\bibfnamefont{B.}~\bibnamefont{Wilhelmi}},
  \bibinfo{journal}{J. Opt. Soc. Am. B}
  \textbf{\bibinfo{volume}{2}}(\bibinfo{number}{4}), \bibinfo{pages}{680}
  (\bibinfo{year}{1985}).

\bibitem[{\citenamefont{Trebino}(2000)}]{Trebino2000}
\bibinfo{author}{\bibfnamefont{R.}~\bibnamefont{Trebino}},
  \emph{\bibinfo{title}{Frequency-Resolved Optical Gating: The Measurement of
  Ultrashort Laser Pulses}}
  (\bibinfo{publisher}{Springer US}, \bibinfo{year}{2000}).

\bibitem[{\citenamefont{Chilla and Martinez}(1991)}]{Chilla1991}
\bibinfo{author}{\bibfnamefont{J.~L.~A.} \bibnamefont{Chilla}}
  \bibnamefont{and} \bibinfo{author}{\bibfnamefont{O.~E.}
  \bibnamefont{Martinez}}, \bibinfo{journal}{Opt. Lett.}
  \textbf{\bibinfo{volume}{16}}(\bibinfo{number}{1}), \bibinfo{pages}{39}
  (\bibinfo{year}{1991}).

\bibitem[{\citenamefont{Iaconis and Walmsley}(1998)}]{Iaconis1998}
\bibinfo{author}{\bibfnamefont{C.}~\bibnamefont{Iaconis}} \bibnamefont{and}
  \bibinfo{author}{\bibfnamefont{I.}~\bibnamefont{Walmsley}},
  \bibinfo{journal}{Opt. Lett.}
  \textbf{\bibinfo{volume}{23}}(\bibinfo{number}{10}), \bibinfo{pages}{792}
  (\bibinfo{year}{1998}).

\bibitem[{\citenamefont{Nicholson} \emph{et~al.}(1999)\citenamefont{Nicholson,
  Jasapara, Rudolph, Omenetto, and Taylor}}]{Nicholson_et_al1999}
\bibinfo{author}{\bibfnamefont{J.~W.} \bibnamefont{Nicholson}},
  \bibinfo{author}{\bibfnamefont{J.}~\bibnamefont{Jasapara}},
  \bibinfo{author}{\bibfnamefont{W.}~\bibnamefont{Rudolph}},
  \bibinfo{author}{\bibfnamefont{F.~G.} \bibnamefont{Omenetto}},
  \bibnamefont{and} \bibinfo{author}{\bibfnamefont{A.~J.}
  \bibnamefont{Taylor}}, \bibinfo{journal}{Opt. Lett.}
  \textbf{\bibinfo{volume}{24}}(\bibinfo{number}{23}), \bibinfo{pages}{1774}
  (\bibinfo{year}{1999}).

\bibitem[{\citenamefont{Xu} \emph{et~al.}(2006)\citenamefont{Xu, Gunn, Cruz,
  Lozovoy, and Dantus}}]{Xu_et_al2006}
\bibinfo{author}{\bibfnamefont{B.}~\bibnamefont{Xu}},
  \bibinfo{author}{\bibfnamefont{J.~M.} \bibnamefont{Gunn}},
  \bibinfo{author}{\bibfnamefont{J.~M.~D.} \bibnamefont{Cruz}},
  \bibinfo{author}{\bibfnamefont{V.~V.} \bibnamefont{Lozovoy}},
  \bibnamefont{and} \bibinfo{author}{\bibfnamefont{M.}~\bibnamefont{Dantus}},
  \bibinfo{journal}{J. Opt. Soc. Am. B}
  \textbf{\bibinfo{volume}{23}}(\bibinfo{number}{4}), \bibinfo{pages}{750}
  (\bibinfo{year}{2006}).

\bibitem[{\citenamefont{Hoppe}(1969)}]{Hoppe1969}
\bibinfo{author}{\bibfnamefont{W.}~\bibnamefont{Hoppe}}, \bibinfo{journal}{Acta
  Crystallographica Section A}
  \textbf{\bibinfo{volume}{25}}(\bibinfo{number}{4}), \bibinfo{pages}{495}
  (\bibinfo{year}{1969}).

\bibitem[{\citenamefont{Rodenburg} \emph{et~al.}(2007)\citenamefont{Rodenburg,
  Hurst, and Cullis}}]{Rodenburg_et_al2007}
\bibinfo{author}{\bibfnamefont{J.~M.} \bibnamefont{Rodenburg}},
  \bibinfo{author}{\bibfnamefont{A.~C.} \bibnamefont{Hurst}}, \bibnamefont{and}
  \bibinfo{author}{\bibfnamefont{A.~G.} \bibnamefont{Cullis}},
  \bibinfo{journal}{Ultramicroscopy} \textbf{\bibinfo{volume}{107}},
  \bibinfo{pages}{227} (\bibinfo{year}{2007}).

\bibitem[{\citenamefont{McCallum and Rodenburg}(1992)}]{McCallumRodenburg1992}
\bibinfo{author}{\bibfnamefont{B.~C.} \bibnamefont{McCallum}} \bibnamefont{and}
  \bibinfo{author}{\bibfnamefont{J.~M.} \bibnamefont{Rodenburg}},
  \bibinfo{journal}{Ultramicroscopy} \textbf{\bibinfo{volume}{45}},
  \bibinfo{pages}{371} (\bibinfo{year}{1992}).

\bibitem[{\citenamefont{Thibault} \emph{et~al.}(2009)\citenamefont{Thibault,
  Dierolf, Bunk, Menzel, and Pfeiffer}}]{Thibault_et_al2009}
\bibinfo{author}{\bibfnamefont{P.}~\bibnamefont{Thibault}},
  \bibinfo{author}{\bibfnamefont{M.}~\bibnamefont{Dierolf}},
  \bibinfo{author}{\bibfnamefont{O.}~\bibnamefont{Bunk}},
  \bibinfo{author}{\bibfnamefont{A.}~\bibnamefont{Menzel}}, \bibnamefont{and}
  \bibinfo{author}{\bibfnamefont{F.}~\bibnamefont{Pfeiffer}},
  \bibinfo{journal}{Ultramicroscopy}
  \textbf{\bibinfo{volume}{109}}(\bibinfo{number}{4}), \bibinfo{pages}{338 }
  (\bibinfo{year}{2009}).

\bibitem[{\citenamefont{Maiden and Rodenburg}(2009)}]{MaidenRodenburg2009}
\bibinfo{author}{\bibfnamefont{A.~M.} \bibnamefont{Maiden}} \bibnamefont{and}
  \bibinfo{author}{\bibfnamefont{J.~M.} \bibnamefont{Rodenburg}},
  \bibinfo{journal}{Ultramicroscopy} \textbf{\bibinfo{volume}{109}},
  \bibinfo{pages}{1256} (\bibinfo{year}{2009}).

\bibitem[{\citenamefont{Guizar-Sicairos}(2010)}]{Guizar-Sicairos2010}
\bibinfo{author}{\bibfnamefont{M.} \bibnamefont{Guizar-Sicairos}} 
  \bibinfo{author}{\bibfnamefont{K.} \bibnamefont{Evans-Lutterodt}},
  \bibinfo{author}{\bibfnamefont{A.~F.} \bibnamefont{Isakovic}},
  \bibinfo{author}{\bibfnamefont{A.} \bibnamefont{Stein}},
  \bibinfo{author}{\bibfnamefont{J.~B.} \bibnamefont{Warren}},
  \bibinfo{author}{\bibfnamefont{A.~R.} \bibnamefont{Sandy}},
  \bibinfo{author}{\bibfnamefont{S.} \bibnamefont{Narayanan}} \bibnamefont{and}
  \bibinfo{author}{\bibfnamefont{J.~R.} \bibnamefont{Fienup}},
  \bibinfo{journal}{Optics Expr.} \textbf{\bibinfo{volume}{18}},
  \bibinfo{pages}{18374} (\bibinfo{year}{2010}).

\bibitem[{\citenamefont{Spangenberg}
  \emph{et~al.}(2014)\citenamefont{Spangenberg, Neethling, Rohwer, Br\"ugmann,
  and Feurer}}]{Spangenberg_et_al2014a}
\bibinfo{author}{\bibfnamefont{D.}~\bibnamefont{Spangenberg}},
  \bibinfo{author}{\bibfnamefont{P.}~\bibnamefont{Neethling}},
  \bibinfo{author}{\bibfnamefont{E.}~\bibnamefont{Rohwer}},
  \bibinfo{author}{\bibfnamefont{M.~H.} \bibnamefont{Br\"ugmann}},
  \bibnamefont{and} \bibinfo{author}{\bibfnamefont{T.}~\bibnamefont{Feurer}},
  \bibinfo{journal}{Phys. Rev. A} \textbf{\bibinfo{volume}{91}} \bibinfo{pages}{021803(R)}  
(\bibinfo{year}{2014}).

\bibitem[{\citenamefont{Faulkner and Rodenburg}(2005)}]{Faulkner2005}
\bibinfo{author}{\bibfnamefont{H.}~\bibnamefont{Faulkner}} \bibnamefont{and}
  \bibinfo{author}{\bibfnamefont{J.}~\bibnamefont{Rodenburg}},
  \bibinfo{journal}{Ultramicroscopy}
  \textbf{\bibinfo{volume}{103}}(\bibinfo{number}{2}), \bibinfo{pages}{153}
  (\bibinfo{year}{2005}).

\bibitem[{\citenamefont{Kane}(2008)}]{Kane2008}
\bibinfo{author}{\bibfnamefont{D.}~\bibnamefont{Kane}},
  \bibinfo{journal}{JOSA B}
  \textbf{\bibinfo{volume}{25}}(\bibinfo{number}{6}), \bibinfo{pages}{A120}
  (\bibinfo{year}{2008}).

\bibitem[{\citenamefont{DeLong}(1995)}]{DeLong1995}
\bibinfo{author}{\bibfnamefont{K.W.}~\bibnamefont{DeLong}},
\bibinfo{author}{\bibfnamefont{R.}~\bibnamefont{Trebino}} \bibnamefont{and}
  \bibinfo{author}{\bibfnamefont{W.E.}~\bibnamefont{White}},
  \bibinfo{journal}{JOSA B}
  \textbf{\bibinfo{volume}{12}}(\bibinfo{number}{12}), \bibinfo{pages}{2463}
  (\bibinfo{year}{1995}).

\bibitem[{\citenamefont{Hornung} \emph{et~al.}(2004)\citenamefont{Hornung,
  Vaughan, Feurer, and Nelson}}]{Vaughan2004}
\bibinfo{author}{\bibfnamefont{T.}~\bibnamefont{Hornung}},
  \bibinfo{author}{\bibfnamefont{J.~C.}~\bibnamefont{Vaughan}},
  \bibinfo{author}{\bibfnamefont{T.}~\bibnamefont{Feurer}}, \bibnamefont{and}
  \bibinfo{author}{\bibfnamefont{K.~A.} \bibnamefont{Nelson}},
  \bibinfo{journal}{Opt. Lett.}
  \textbf{\bibinfo{volume}{29}}(\bibinfo{number}{17}), \bibinfo{pages}{2052}
  (\bibinfo{year}{2004}).

\end{thebibliography}
\end{document}